\documentclass[twocolumn]{aastex631}
\usepackage{braket}

\shorttitle{MARs of the HectoMAP galaxy clusters}
\shortauthors{Pizzardo et al.}

\begin{document}

\title{Mass accretion rates of the HectoMAP clusters of galaxies}

\author{M. Pizzardo}
\affiliation{Dipartimento di Fisica, Universit\`a degli Studi di Torino, via P. Giuria 1,  I-10125 Torino, Italy }
\affiliation{Istituto Nazionale di Fisica Nucleare (INFN), Sezione di Torino, via P. Giuria 1,  I-10125 Torino, Italy}

\author{J. Sohn}
\affiliation{Smithsonian Astrophysical Observatory, 60 Garden Street, Cambridge, MA-02138, USA}

\author{M. J. Geller }
\affiliation{Smithsonian Astrophysical Observatory, 60 Garden Street, Cambridge, MA-02138, USA}

\author{A. Diaferio}
\affiliation{Dipartimento di Fisica, Universit\`a degli Studi di Torino, via P. Giuria 1,  I-10125 Torino, Italy }
\affiliation{Istituto Nazionale di Fisica Nucleare (INFN), Sezione di Torino, via P. Giuria 1,  I-10125 Torino, Italy}

\author{K. Rines}
\affiliation{Department of Physics and Astronomy, Western Washington University, Bellingham, WA-98225, USA}

\begin{abstract}
We estimate the mass accretion rate (MAR) of 321 clusters of galaxies in the HectoMAP Cluster Survey.
The clusters span the redshift range $0.17-0.42$ and the $M_{200}$ mass range $\approx (0.5 - 3.5)\cdot 10^{14} M_\odot$. The MAR estimate is based on the caustic technique along with a spherical infall model. 
Our analysis extends the measurement of MARs for 129 clusters at $z<0.3$ from the Cluster Infall Regions in the Sloan Digital Sky Survey and the Hectospec Cluster Survey to redshift
$z \sim 0.42$. 
Averaging over redshift, low-mass clusters with masses near $0.7\cdot 10^{14} M_\odot$ roughly accrete $3\cdot 10^4 M_\odot$~yr$^{-1}$; more massive clusters with masses near $2.8\cdot 10^{14} M_\odot$ roughly accrete $ 1\cdot 10^5 M_\odot$yr$^{-1}$.
Low- and high-mass clusters increase their MAR by approximately  $46\%$ and $84\%$, respectively, as the redshift increases from  $z$ in the range $0.17-0.29$ to $z$ in the range $0.34-0.42$.
The MARs at fixed redshift increase with mass and  MARs at fixed mass increase with redshift in agreement with the $\Lambda$CDM cosmological model for hierarchical structure formation. We
consider the extension of MAR measurements to $z \sim 1$.
\end{abstract}


\section{Introduction}

In the standard  $\Lambda$CDM model for cosmological structure formation, the laws of gravity follow general relativity in a flat space-time with a positive cosmological constant, $\Lambda$.  Collisionless cold dark matter (CDM) dominates the matter content of the universe. In this model, cosmological structures form hierarchically \citep[e.g.,][]{press1974formation,white1978,bower1991,laceyCole93,sheth2002,zhang2008,corasaniti2011,desimone2011,achitouv2014,musso2018}.             

The outskirts of clusters of galaxies are  a potentially powerful probe of the model of structure formation and evolution \citep[e.g.,][]{diaferio2004outskirts,reiprich2013outskirts,Diemer2014,lau2015mass,walker2019physics,rost2021threehundred}: 
Far from the center (at radii $\gtrsim 2R_{\rm vir}$\footnote{ $R_{\mathrm {vir}}$ is the radius within which the cluster dynamics satisfies the virial theorem. This radius is usually estimated by assuming spherical top-hat collapse of a density perturbation. $R_{\Delta}$ is commonly defined as the radius enclosing an  average mass density  $\Delta$ times the critical density of the Universe at the appropriate redshift. The value $\Delta=200$ thus defines $R_{200}$. For a $\Lambda$CDM model with cosmic mass density $\Omega_m=0.3$ and cosmological constant $\Omega_\Lambda=0.7$, $R_{\mathrm {vir}}\sim R_{100}$ \citep{bryan1998}.}), hydrodynamic simulations show that baryonic processes play a minor role in cluster dynamics \citep[e.g.,][]{diemand2004,vandaalen2014,velliscig2014,hellwing2016,armitage2018,shirasaki2018}. This feature makes the comparison with simulations more robust because the relevant physical processes are dominated by gravity. 

The mass accretion rate (MAR) of galaxy clusters offers a valuable tool for probing the outskirts of clusters. $N$-body simulations show that the infall region of clusters, where new material is falling onto the clusters for the first time, is located beyond $\sim 2R_{200}$ \citep{ludlow2009,Diemer2014,More2015,Diemer2017sparta2,bakels2020,Xhakaj2020}. 
The MAR is tightly linked to the cluster properties \citep{Vitvitska_2002,gao2004sub,
vandenbosch2005,kasun2005shapes,allgood2006shape,bett2007spin,ragone2010relation,Ludlow2013,Diemer2014,More2015,PizzardoPrep}. For example, consider the splashback radius, the average location of the first apocentre of infalling material \citep{adhikari2014}. At the splashback radius, there is a sudden decrease of the radial logarithmic derivative of the density profile, ${\mathrm d} \log \rho / {\mathrm d} \log r$ \citep{Diemer2014,More2015}. $N$-body simulations show that the location of the splashback radius depends on the cluster mass and on the MAR: larger masses and accretion rates are associated with smaller splashback radii \citep{Diemer2014,More2015,PizzardoPrep}.

Numerous theoretical studies investigate the mass accretion history (MAH), $M(z)$, and the related MAR, $\dot{M}(z)$, in $\Lambda$CDM cosmologies. Generally, these studies use numerical simulations to build merger trees for samples of halos with different evolved masses at $z=0$, $M_0$. The simulations trace the cluster accretion back in time by identifying their main massive progenitors at increasing redshift. 
Different phenomenological analytical functions \citep{vandenbosch02,mcbride2009,Fakhouri2010,Correa15,Diemer2017sparta2},
or semianalytical models \citep[e.g.,][]{vandenbosch14},  based on the extended Press-Schecther (EPS) formalism \citep{Bond1991}, describe these results.
In all of these simulations, halos of evolved mass $M_0 \sim (0.1-5)\cdot 10^{14} M_\odot$ accrete $\sim 30\%-50\%$ of their mass from $z\sim 0.5$ to $z\sim 0$. The MAR increases with $M_0$: in $z\sim 0-0.5$, halos with $M_0 \sim 3.5\cdot 10^{13} M_\odot$ have MAR~$\sim 3\cdot 10^3 M_\odot$~yr$^{-1}$, whereas halos with $M_0 \sim 10^{14} M_\odot$ have MAR~$\sim 10^4 M_\odot$~yr$^{-1}$.

Despite the importance of the cluster outskirts, observations are limited compared with the intensive study of clusters within their approximate virial radii, $R_{200}$. There are two reasons for this contrast. Because these regions cover a  large region on the sky and the number density of galaxies in the outer regions of clusters is much smaller than in the inner regions, the contrast with the foreground/background is much lower, and it is challenging to obtain substantial samples of cluster members at these large radii. Other observational techniques for estimating  mass profiles at large cluster-centric distances are also limited. The X-ray surface brightness drops at large radii. Sunyaev-Zel'dovich (SZ) effect measurements are  less sensitive at large radii because the gas pressure is low. The weak-gravitational-lensing signal is more sensitive to projection effects at large radius \citep{Serra2011,Ettori2013rev,reiprich2013outskirts}.

Most techniques for estimating the mass profiles of clusters assume dynamical equilibrium \citep{zwicky1937masses,the1986jeans,merritt1987,sarazin1988x,pierpaoli2003,rasia2006,Serra2011,Ettori2013rev,reiprich2013outskirts}. 
Only weak gravitational lensing  \citep{bartelmann2010gravitational,hoekstra2013masses,Umetsu20essay} and the caustic technique \citep{Diaferio1997,Diaferio99,Serra2011} avoid the assumption of dynamical equilibrium. Both techniques can be applied at larger radii. Weak-lensing estimates are also redshift dependent, with a signal that peaks at intermediate redshift \citep{Hoekstra2003,Hoekstra2011}. Both methods are steadily improving with the acquisition of large combined spectroscopic and photometric datasets.  There are also  advances in lensing-mass reconstruction methods.
\citet{Umetsu11} and \citet{Umetsu13} improve the precision of cluster-mass measurements from weak lensing by $\sim 30\%$ with the joint use of lensing distortion and magnification.
Using this method on the CLASH survey \citep{Postman2012clash}, \citet{umetsu2014clash} estimate the clusters' mass profiles for radii $\lesssim 3$~Mpc. \citet{Umetsu16} include strong gravitational lensing shear and magnification and extend the estimates to even larger distances, $\sim 5.7$~Mpc.

Although the caustic technique is also subject to projection effects, it returns an unbiased mass estimate with a relative uncertainty of 50\% at large radii for sufficiently densely sampled systems. The caustic technique is independent of redshift \citep{Serra2011}, but obtaining large samples of cluster members at higher redshift is more demanding. In principle, weak lensing and the caustic technique offer complementary approaches that combine to elucidate the physics of the outer regions of clusters. Here we focus our attention on the application of the caustic technique. 
  
Taking advantage of the Cluster Infall Regions in the Sloan Digital Sky Survey (CIRS) \citep{Rines2006CIRS} and the Hectospec Cluster Survey (HeCS) \citep{Rines2013HeCS} spectroscopic surveys (now included in the HeCS-omnibus; \citet{sohn2019velocity}), \citet{pizzardo2020} first estimated the MARs of 129 clusters in the redshift range $0.01<z<0.3$ and mass range $M_{200}\sim (0.1-25)\cdot 10^{14} M_\odot$. They adopt a  procedure based on the caustic technique to estimate the mass profile of the clusters, and they apply a spherical infall model \citep{deBoni2016} to estimate the cluster MARs. The MAR estimation method is informed by $N$-body simulations. 

The resultant  MARs increase with mass at fixed redshift  and with redshift at fixed mass in agreement with $\Lambda$CDM predictions \citep[see][for further information]{pizzardo2020}. Clusters with masses $\sim 1.6 \cdot 10^{14} M_\odot$ have MARs $\sim 3\cdot 10^4 M_\odot$ yr$^{-1}$, clusters with masses $\sim 4.4 \cdot 10^{14} M_\odot$ have MARs $\sim 8\cdot 10^4 M_\odot$ yr$^{-1}$; clusters with considerably higher masses, $\sim 1 \cdot 10^{15} M_\odot$, have MARs $\sim 1.6\cdot 10^5 M_\odot$ yr$^{-1}$. These results are in essential agreement with the MARs of dark matter halos estimated in $\Lambda$CDM  $N$-body simulations \citep{vandenbosch02,mcbride2009,Fakhouri2010,vandenbosch14,Diemer2017sparta2}.

The MAR estimates of \citet{pizzardo2020} are limited to redshifts $z \lesssim0.3$. Extension to higher redshift offers the opportunity for  more sensitive tests of models for the growth of structure in the universe. We thus explore the additional constraints obtained from an independent cluster sample reaching $z \sim 0.42$. \citet{sohn2021cluster} identified 346 clusters with redshift $0.17\lesssim z\lesssim 0.42$ by applying a Friends-of-Friends (FoF) algorithm to the full HectoMAP redshift survey \citep{Geller11, Hwang16}. The HectoMAP survey includes $\sim 110,000$ galaxies with spectroscopic redshifts with $z<0.6$ and with an average redshift $\hat{z} \sim 0.31$ \citep[see][for the first data release]{sohn2021hectomap}.

In Section \ref{sec:hectomap-pres} we review the HectoMAP  cluster sample. In Section \ref{sec:hectomap-mar} we briefly introduce the recipe we use for  estimating of the MAR.  We then measure the MARs of the HectoMAP clusters. In Section \ref{sec:discussion} we discuss the results. We conclude in Section \ref{sec:conclusion}. We use the standard $\Lambda$CDM parameters $\Omega_{m0}=0.27$, $\Omega_{\Lambda 0}=0.73$, and $H_0=70$~km~s$^{-1}$~Mpc$^{-1}$.

\section{HectoMAP Cluster Sample}
\label{sec:hectomap-pres}

HectoMAP is a dense redshift survey designed to study galaxy clustering at redshifts $0.2 < z < 0.6$ (\citealp{Geller11, Hwang16, sohn2021hectomap}). \citet{sohn2021cluster} use an FoF algorithm to identify galaxy clusters in HectoMAP. \citet{sohn2021hectomap} and \citet{sohn2021cluster} describe the details of the HectoMAP redshift survey and the HectoMAP FoF cluster catalog. We briefly review the HectoMAP survey in Section \ref{sec:hmap} and the HectoMAP FoF cluster catalog in Section \ref{sec:fof}. 

\subsection{The HectoMAP Redshift Survey}\label{sec:hmap}

HectoMAP is a large-scale redshift survey covering 54.64 deg$^{2}$ over a narrow strip of the sky at $200 <$ R.A. (deg) $< 250$ and $42.5 <$ Decl. (deg) $< 44.0$. HectoMAP is based on Sloan Digital Sky Survey (SDSS) Data Release (DR) 16 photometry \citep{Ahumada20}. The HectoMAP redshift survey includes a modest number of redshifts from SDSS DR16. \citet{sohn2021hectomap} measured most of the redshifts with the Hectospec instrument mounted on the MMT 6.5 m telescope \citep{Fabricant98, Fabricant2005Hecto}. 

The primary targets of HectoMAP are galaxies with $r < 20.5$ and $(g-r) > 1$, and galaxies with $20.5 \leq r < 21.3$, $g-r > 1$, and $r-i>0.5$. The full survey is $>80\%$ complete at $r = 20.5$ and 62\% complete at $r = 21.3$. The completeness of the survey is much less for bluer galaxies outside the target color range \citep{sohn2021hectomap}. For red objects within the selection limits, the HectoMAP survey has remarkably uniform completeness on the sky. This uniformity results from the strategy of revisiting every position in the field typically $\sim 9$ times (see Section 2.4 of \citealt{sohn2021hectomap} for details). The uniformity of HectoMAP makes the survey a robust basis for the examination of properties of dense systems and their evolution. 

HectoMAP includes a total of  $\sim 110,000$ galaxies with spectroscopic redshifts; the average galaxy number density is $\sim 2000$ galaxies deg$^{-2}$. The typical uncertainty of HectoMAP redshifts is $\sim 40$ km s$^{-1}$. The high density and uniform completeness of the survey enable the identification and study of HectoMAP galaxy clusters based on spectroscopy \citep{sohn2021cluster}. 

\subsection{HectoMAP FoF Cluster Catalog}\label{sec:fof}

\citet{sohn2021cluster} built a cluster catalog by applying an FoF algorithm to the HectoMAP redshift survey. The FoF algorithm bundles sets of neighboring galaxies within given linking lengths into candidate clusters. \citet{sohn2021cluster} apply the FoF algorithm in redshift space. The standard FoF algorithm requires two linking lengths: one in the projected direction ($\Delta D$) and one in the radial direction ($\Delta V$). 

\citet{sohn2021cluster} choose the optimal linking lengths empirically using photometrically identified redMaPPer clusters \citep{Rykoff2014,Rykoff2016}. The redMaPPer cluster catalog includes 104 redMaPPer clusters in HectoMAP. \citet{sohn2021cluster} demonstrate that the FoF algorithm with $\Delta D = 900$ kpc and $\Delta V = 500$ km s$^{-1}$ identifies more than 90\% of the HectoMAP redMaPPer clusters. With these linking lengths, the FoF algorithm also identifies all of the 15 known X-ray clusters in HectoMAP \citep{Sohn2018b}. 

The HectoMAP FoF cluster catalog includes 346 systems with 10 or more spectroscopic members. \citet{sohn2021cluster} report properties of the FoF clusters including the number of FoF members, the cluster center, and the cluster velocity dispersion. The HectoMAP FoF clusters typically consist of 17 members. 

We use the center of each cluster as listed in \citet{sohn2021cluster}. The centers of the FoF clusters are determined based on the center-of-light method \citep{Robotham11}, which weights the positions of FoF members by their luminosity. This method identifies a bright central galaxy. \citet{sohn2021cluster} identify the position and redshift of this central galaxy as the  FoF cluster center. Finally, \citet{sohn2021cluster}  use the biweight technique \citep{Beers1990} to compute the line-of-sight velocity dispersion of the cluster. 

We select spectroscopic galaxies within $R_{\rm cl} < 5$ Mpc and $|\Delta cz / (1+z_{\rm cl})| < 5000~{\rm km~s}^{-1}$ of each cluster center as a basis for measuring the mass accretion rate. The line-of-sight velocity boundaries are $\gtrsim 5$ times larger than the velocity dispersions of HectoMAP clusters. The radial boundaries are much larger than the region where we measure the mass accretion rate (see Figure \ref{r_vlos_stk}, and Tables \ref{stkBins} and \ref{rm200_mar_stk}). Thus, the selection of the survey boundaries does not impact our analysis. 

\section{The MAR of HectoMAP Clusters}
\label{sec:hectomap-mar}
The HectoMAP clusters span an interesting redshift range where clusters are expected to accrete roughly half of their mass \citep{vandenbosch02,mcbride2009,Fakhouri2010,vandenbosch14}. We estimate the MARs of the HectoMAP clusters for comparison with model predictions.

In Section \ref{sec:recipe} we briefly review the procedure for estimating the MAR from spectroscopic catalogs \citep{pizzardo2020}. In Section \ref{sec:ct-profiles} we describe the sample of stacked HectoMAP clusters and their mass profiles. In Section \ref{sec:mar-estimates} we estimate the MAR of the HectoMAP clusters and compare the results with previous measurements and with  $\Lambda$CDM predictions.

\subsection{The MAR Recipe}\label{sec:recipe}

\citet{pizzardo2020} use the MAR definition proposed by \citet{deBoni2016}. This approach assumes that a spherical infall model describes the accretion of new material onto a cluster. During the infall time $t_{\rm inf}$, a cluster accretes all of the material within a spherical shell centered on the cluster center and with an inner radius $R_{i}$ and thickness $\delta_s$. 

We derive the shell thickness from the classical equation of motion for accretion with constant acceleration and initial velocity $v_{\rm inf}$,
\begin{equation}\label{eq:thickness}
t_{\rm inf}^2  GM(<R_{i}) - t_{\rm inf} 2 R_{i}^{2} (1 + \delta_{s}/2)^2 v_{\rm inf} - R_{i}^{3}  \delta_{s} (1 + \delta_{s}/2)^{2} = 0,
\end{equation}
where $M (< R_{i})$ is the mass of the cluster within a radius $R_{i}$, and $G$ is the gravitational constant.

Given the mass of the infalling shell, $M_{\rm shell}$, the estimate of the MAR is 
\begin{equation}\label{mar}
{\rm MAR} \equiv \frac{M_{\rm shell}}{t_{\rm inf}}. 
\end{equation}
\citet{deBoni2016} and \citet{pizzardo2020} set $t_{\rm inf}=1$ Gyr and $R_{i}=2 R_{200}$. To estimate the infall velocity, $v_{\rm inf}$, they rely on radial velocity profiles of simulated $\Lambda$CDM halos. For bins in redshift and mass, they compute this profile based on the set of median values of the radial velocities for all of the particles in each of 200 radial bins in the range $(0,10)~R_{200}$. The particle radial velocity is $\mathrm{v}_{i} = [{\mathbf v}_p +H(z_s)a(z_{s}){\mathbf r}_{c,i}]\cdot {\mathbf r}_{c,i}/r_{c,i}$, where ${\mathbf v}_{p}$ is the proper peculiar velocity and ${\mathbf r}_{c,i}$ is the comoving position vector of the particle relative to the cluster center. At the snapshot redshift $z_s$, $H(z_{s})$ and $a(z_{s})$ are the Hubble parameter and the cosmic scale factor.  

The minimum of the velocity profile is generally within the radial range $[2,2.5] R_{200}$. This range thus provides the most reliable estimate of the MAR. We adopt $v_{\rm inf}$ as the value of the infall velocity for the median profile in the radial range $[2,2.5]~ R_{200}$.

The radial position of $v_{\rm inf}$, and thus of the infall region of a cluster, is coincident with the approximate location of the splashback radius \citep{Diemer2014,More2015}.  $N$-body simulations show that the splashback radius separates the inner region where accretion is nearly complete from the outer region where accretion is ongoing \citep{Diemer2017sparta2,Xhakaj2020}.

Defining the observable MAR relies on the $v_{\rm inf}$ derived from $N$-body simulations. This approach is necessary because measurement of the radial velocity profiles of observed cluster galaxies is not currently feasible. The results are insensitive to the exact choice of $v_{\rm inf}$. For example, \citet{pizzardo2020} show that varying $v_{\rm inf}$ within $\pm 40\%$ of the true value produces results within the typical $\sim 40\%$ spread of the resulting MARs.

\subsection{Mass Profiles from the Caustic Technique}\label{sec:ct-profiles}
\subsubsection{The Caustic Technique}\label{sec:ct}

Use of Eqs. (\ref{eq:thickness}) and (\ref{mar}) requires an estimate of the cluster mass at large cluster-centric distances where virial equilibrium does not hold. \citet{pizzardo2020} use the caustic method \citep{Diaferio1997,Diaferio99,Serra2011}. At large distances, $(0.6-4)R_{200}$, the caustic method returns an unbiased estimate of the mass with better than 10\% accuracy and with a relative uncertainty of $50\%$ \citep{Serra2011} provided that the velocity field of the cluster outer region is sufficiently well sampled. According to \citet{Serra2011}, $\sim 200$ spectroscopic measurements within a three-dimensional distance $\sim 3 R_{200}$ provide a sufficient sample. Increased sampling generally improves performance. Significantly sparser samples lead to underestimation of the mass. For example, a sample of only $\sim100$ spectroscopically identified members results in a mass underestimated by $\sim 35\%$.

\citet{pizzardo2020} demonstrate that, in $\Lambda$CDM simulations, the MARs estimated by applying the spherical accretion recipe based on caustic mass profiles are unbiased. The MARs are within $\sim 19\%$ of the MARs computed by applying the same accretion recipe to the true mass profiles of large samples of simulated clusters.

The caustic technique estimates the three-dimensional mass profile of a cluster from the line-of-sight escape velocity of the cluster galaxies as a function of cluster-centric distance $R$, $\braket{v_{\rm esc, los}^2}(R)$. The $R-v_{\rm los}$ diagram, the line-of-sight velocity relative to the cluster median as a function of $R$, is the basis for the extraction of the caustic mass profile. In this diagram, the cluster galaxies appear in a well-defined trumpet-shaped pattern; the amplitude decreases as $R$ increases. The vertical separation between the upper and lower caustics at radius $R$ is the caustic amplitude, $\mathcal{A}(R)$, which approximates the average line-of-sight escape velocity profile. The caustic technique locates the caustics from the $R-v_{\rm los}$ diagram. The square of the caustic amplitude, $\mathcal{A}^2(R)$, estimates $\braket{v_{\rm esc, los}^2}(R)$. 
With a form (or filling) factor $\mathcal{F}_\beta$ accounting for the cluster velocity anisotropy,\footnote{  $\mathcal{F}_\beta$ is the average of a function that combines the profiles of the mass density $\rho(r)$ and of the  velocity anisotropy parameter $\beta(r)$ of the cluster. In hierarchical clustering scenarios, this function is weakly dependent on the cluster-centric distance $r$ at $r\gtrsim R_{200}$ and can be replaced by its average $\mathcal{F}_\beta$ (see \citealt{Diaferio99} and \citealt{Serra2011} for further details). } 
 the caustic technique provides an estimate of the three-dimensional escape velocity profile, $\braket{v_{\rm esc}^2}(R)$, which is related to the three-dimensional cluster gravitational potential $\phi$, $\braket{v_{\rm esc}^2}(R)=-2\phi$. The resulting estimate of the mass profile is 
\begin{equation}\label{eq:mass-prof}
G M(<R) = \mathcal{F}_\beta \int_0^R \mathcal{A}^2(r)\,dr,
\end{equation}
where $G$ is the gravitational constant and $\mathcal{F}_\beta=0.7$ \citep[see][for further details]{Diaferio1997,Diaferio99,Serra2011}.

\subsubsection{Stacking the Clusters}
\label{sec:stk-prof}

A sufficiently dense spectroscopic catalog is a fundamental requirement for the application of the caustic technique. The HectoMAP cluster catalogs include a median of 102 member galaxies within a projected cluster-centric distance of $5~$Mpc, with a $68$th percentile range (65--157). For these clusters, $R_{200}\sim 1~$Mpc and thus the HectoMAP catalogs include a median of $\sim 50$ galaxies within a projected distance of $3 R_{200}$. This population is an upper limit to the number of galaxies within the three-dimensional cluster-centric distance, $\sim 3 R_{200}$. In general, individual HectoMAP systems are not sufficiently well sampled for optimal performance of the caustic technique.

Given the limitations of redshift sampling for individual HectoMAP clusters, we estimate the MAR of HectoMAP clusters as a function of their velocity dispersion and redshift by stacking the observed clusters. We construct 10 stacked samples in five redshift bins, in the range $[0,0.6]$, and two velocity dispersion bins, $[200,400)$~km~s$^{-1}$ and $[400,1020]$~km~s$^{-1}$. We select the FoF clusters to build these samples based on the velocity dispersion of  each cluster (\citealp{sohn2021cluster}, see Section \ref{sec:fof}). Our procedure removes 25 FoF groups with velocity dispersion $ < 200$~km~s$^{-1}$. These systems have masses $\lesssim 10^{13} M_\odot$. The stacked clusters include 321 individual FoF clusters.

We compute the velocity dispersion of each ensemble cluster by applying the biweight technique \citep{Beers1990} to the ensemble catalog of relative line-of-sight velocities of  galaxies with respect to the central galaxy of the individual FoF cluster. In this computation, we identify the mean line-of-sight velocity of a constituent FoF cluster with the redshift of its central galaxy. We use the bootstrap technique to derive the uncertainty in the  ensemble velocity dispersion. Table \ref{stkBins} lists the total number of galaxies, the median redshift and its $68$th percentile range, and the velocity dispersion for each of the stacked clusters.
\begin{table*}[htbp]
\begin{center}
\caption{\label{stkBins} Stacked clusters.}
\begin{tabular}{ccccc}
\hline
\hline
Stacked Cluster & No. of galaxies & Median $z$ & 68th Percentile Range & Bi-weighted $\sigma_{v_{\rm los}}$\\
 &  &  & (redshift $z$) & (km~s$^{-1}$) \\
\hline
 & & & & \\
z1sig1 & 5652 & 0.166 & (0.114-0.211) & 360.5 $\pm$ 9.1 \\ 
z1sig2 & 2427 & 0.181 & (0.135-0.215) & 640 $\pm$ 18 \\
z2sig1 & 4619 & 0.243 & (0.223-0.257) & 356.5 $\pm$ 8.9 \\
z2sig2 & 3747 & 0.244 & (0.226-0.264) & 818 $\pm$ 19 \\
z3sig1 & 4356 & 0.286 & (0.272-0.293) & 356.0 $\pm$ 9.5 \\
z3sig2 & 2797 & 0.288 & (0.273-0.295) & 637 $\pm$ 19 \\
z4sig1 & 3510 & 0.339 & (0.316-0.370) & 358 $\pm$ 11 \\
z4sig2 & 2493 & 0.345 & (0.324-0.371) & 582 $\pm$ 20 \\
z5sig1 & 3146 & 0.417 & (0.382-0.454) & 376 $\pm$ 13 \\
z5sig2 & 2314 & 0.420 & (0.391-0.479) & 609 $\pm$ 20 \\
\hline
\end{tabular}
\end{center}
\end{table*}

We then apply the caustic technique to the 10 stacked clusters in Table \ref{stkBins}. For each ensemble cluster, we build an $R-v_{\rm los}$ diagram containing all of the galaxies with projected distance from their respective host FoF cluster center $< 5~$Mpc and with absolute line-of-sight velocity relative to the host cluster mean $ <5000$~km~s$^{-1}$. We use centers of the individual FoF clusters listed in \citet{sohn2021cluster} who identify the mean FoF cluster redshift with the redshift of the central galaxy. Finally, we derive the caustics for each stacked $R-v_{\rm los}$ diagram and compute the related mass profiles according to Eq. (\ref{eq:mass-prof}). 

\begin{figure*}
\begin{center}
\includegraphics[scale=0.48]{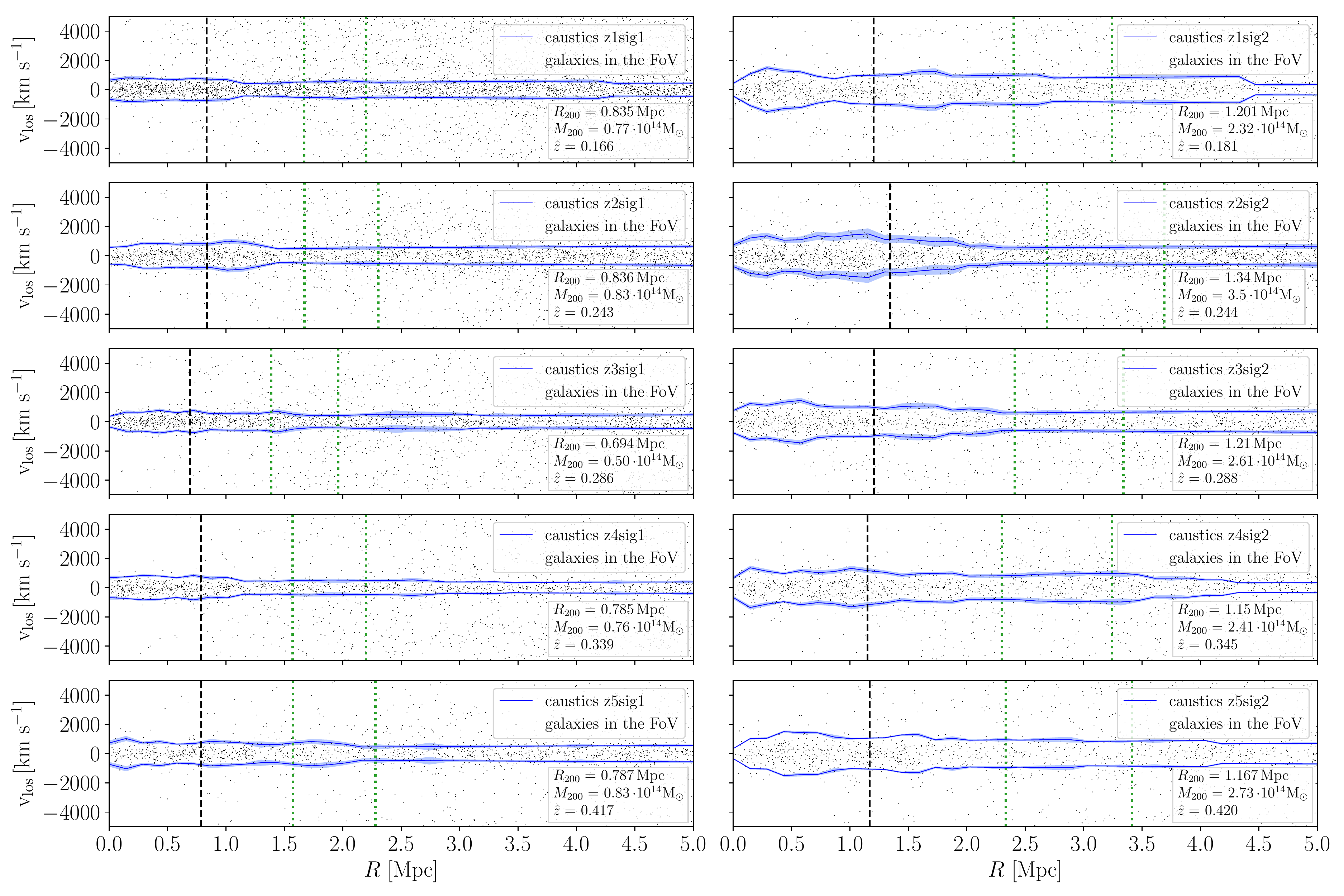}
\caption{$R-v_{\rm los}$ diagrams for the 10 stacked clusters in Table \ref{stkBins}. The blue curves indicate the caustics; the dashed black lines show $R_{200}$ derived from the caustic mass profiles, and the green dotted lines delimit the infall shell for estimation of the MAR according to the \citet{pizzardo2020} procedure.}\label{r_vlos_stk}
\end{center}
\end{figure*}

Following \citet{pizzardo2020}, the $R-v_{\rm los}$ diagrams of the stacked clusters are based on the projected cluster-centric distances and line-of-sight velocities of the galaxies in the reference frame of each of the individual systems included in the stack. In other words, there is no additional normalization. This approach is robust if the bins in velocity dispersion are small enough to avoid the presence of systems with widely different masses. The procedure avoids introducing systematics from the uncertain individual  constituent cluster estimate of $R_{200}$ and velocity dispersion. Adopting a mass-velocity dispersion relation \citep[e.g.,][]{Evrard2008Msigma}, the 90th percentile range of the clusters in the low- and high-velocity dispersion bins contains systems with masses in the range $ (0.1-0.8)\cdot 10^{14} M_\odot$ and $ (1-4)\cdot 10^{14} M_\odot$, respectively. Figure \ref{r_vlos_stk} shows the $R-v_{\rm los}$ diagrams of the 10 stacked clusters of Table \ref{stkBins} and the caustics. Table \ref{rm200_mar_stk} lists the $R_{200}$ and $M_{200}$ for each stacked cluster along with their uncertainties.
\begin{table*}[htbp]
\begin{center}
\caption{\label{rm200_mar_stk} $R_{200}$, $M_{200}$, $v_{\rm inf}$, and MAR of the Stacked Clusters.}
\begin{tabular}{cccccc}
\hline
\hline
Stacked Cluster & Median $z$ & $R_{200}$ & $M_{200}$ & $v_{\rm inf}$ & MAR \\
 & & (Mpc) & ($10^{14} M_\odot$) & (km~s$^{-1}$) & ($10^3 M_\odot$yr$^{-1}$) \\
\hline
& & & & \\
z1sig1 & 0.166 & $0.835\pm 0.078$ & $0.77\pm 0.22$ & $-190.5\pm 1.2$ & $27\pm 12$\\ 
z1sig2 & 0.181 & $1.201\pm 0.088$ & $2.32\pm 0.51$ & $-285.8\pm 5.3$ & $107\pm 29$\\ 
z2sig1 & 0.243 & $0.836\pm 0.084$ & $0.83\pm 0.25$ & $-219.5\pm 1.3$ & $27\pm 14$\\ 
z2sig2 & 0.244 & $1.34\pm 0.22$ & $3.5\pm 1.7$ & $-381.2\pm 8.3$ & $54\pm 26$\\ 
z3sig1 & 0.286 & $0.694\pm 0.068$ & $0.50\pm 0.15$ & $-207.97\pm 0.37$ & $24\pm 14$ \\ 
z3sig2 & 0.288 & $1.21\pm 0.10$ & $2.61\pm 0.67$ & $-341.9\pm 6.2$ & $60\pm 23$\\ 
z4sig1 & 0.339 & $0.785\pm 0.064$ & $0.76\pm 0.19$ & $-234.5\pm 1.1$ & $23\pm 13$\\ 
z4sig2 & 0.345 & $1.15\pm 0.13$ & $2.41\pm 0.83$ & $-344.2\pm 6.2$ & $129\pm 46$\\ 
z5sig1 & 0.417 & $0.787\pm 0.082$ & $0.83\pm 0.26$ & $-255.5\pm 1.4$ & $53\pm 30$\\ 
z5sig2 & 0.420 & $1.167\pm 0.083$ & $2.73\pm 0.58$ & $-395.4\pm 8.1$ & $143\pm 38$\\ 
\hline
\end{tabular}
\end{center}
\end{table*}

This stacking method, adopted by \citet{Rines2006CIRS} and \citet{Serra2011}, is particularly suitable with the HectoMAP FoF catalog because the $R-v_{\rm los}$ diagram of a stacked cluster is generally dense enough to guarantee  good performance of the caustic technique. Furthermore, the caustic technique assumes spherical symmetry for the cluster member distribution, but individual clusters are often triaxial \citep{Frenk1988,Dubinski1991,Warren1992}. Stacking the member galaxies within  multiple clusters averages out the asphericity of individual clusters \citep{Rines2006CIRS,Serra2011}. \citet{pizzardo2020} demonstrate the validity of this procedure by showing that, at comparable mass and redshift, the MARs of the stacked clusters agree with the average MARs of individual systems. They also show that the method is very robust against the overrepresentation of the richest individual systems and that it is insensitive to the impact of fore- and background galaxies.
The red selection of  HectoMAP clusters members (Sect. \ref{sec:hmap}) does not affect our measurements: \citet{pizzardo2020} show that missing $\sim 35\%$ of the blue galaxies has no impact on the MARs.  \citet{Rines2013HeCS} also show that blue galaxies have undetectable effects on dynamical mass estimates.

\subsection{Estimates of the MAR}\label{sec:mar-estimates}

In addition to the caustic mass profiles of the 10 stacked HectoMAP clusters, we also  require the initial radial velocities of the infalling shells to estimate the MAR. To obtain these velocities, we use  numerical simulations (Sect. \ref{sec:recipe}).

Following \citet{pizzardo2020}, we use the $\Lambda$CDM run of the L-CoDECS $N$-body simulations \citep{Baldi2012CoDECS}. The simulation has a box size of $1.43\, ~{\rm{Gpc}} $  in comoving coordinates. The  simulation includes two collisionless fluids of $1024^3$ particles each, with masses $m_{\rm DM} = 8.34 \times 10^{10} M_{\odot} $ and $m_b = 1.67 \times 10^{10} M_{\odot}$, respectively. 
The simulation is normalized at the cosmic microwave background epoch, with cosmological dark matter density $\Omega_{m0}=0.226$, cosmological constant $\Omega_{\Lambda 0}=0.729$, baryonic mass density $\Omega_{b0}=0.0451$, Hubble constant $H_0=70.3$~km~s$^{-1}$~Mpc$^{-1}$, power spectrum normalization $\sigma_8=0.809$, and power spectrum index $n_s=0.966$. An FoF algorithm identifies groups and clusters in the simulations. The system centers are identified as the most bound particle within the system. 

The $\Lambda$CDM run of L-CoDECS is a standard collisionless $N$-body simulation.
For measuring the MAR, the simulation must trace the galaxy velocity field reliably in the outer regions of clusters, beyond $2 R_{200}$: $N$-body/hydrodynamical simulations \citep[e.g.,][]{diemand2004,hellwing2016,armitage2018} show that indeed the velocity bias between the velocity dispersions of galaxies and of dark matter particles is negligible in the outskirts of galaxy clusters. Hence, collisionless $N$-body simulations provide an unbiased measure of the MAR.
  
Here, we consider the same 12 samples of simulated clusters used by \citet[][see their Sect. 3.1 and Table 1]{pizzardo2020}. At $z=0$,  they include two samples of 2000 and 50 halos with median masses $M_{200} \simeq 1.43\times 10^{14} M_{\odot}$ and $M_{200} \simeq 1.43\times 10^{15} M_{\odot}$, respectively. They trace the main progenitors of these halos at five higher redshifts with  $z \leq 0.44$. Over this redshift range, six low-mass bin samples cover the mass range $\sim (0.6-1.6)\cdot 10^{14} M_{\odot}$ and six high-mass bin samples cover the range $\sim (4-16)\cdot 10^{14} M_{\odot}$.

We derive the median radial velocity profiles of the 12 samples of simulated clusters. We then use these radial profiles to compute the radial infall velocity, $v_{\rm inf}$, for each stacked cluster according to its redshift and mass. The procedure involves three interpolations in  velocity-redshift, mass-redshift, and velocity-mass space. We also compute the uncertainty in $v_{\rm inf}$, but we ignore the uncertainty in  $v_{\rm inf}$  when estimating the MAR with Eq. (\ref{mar}) because this uncertainty is negligible compared with the uncertainty in the mass profile. Sect. 3.3 of \citet{pizzardo2020} contains a more detailed description of this procedure.
For each simulated cluster, Table \ref{rm200_mar_stk} lists the infall velocities and their uncertainty.

\begin{figure}
\begin{center}
\includegraphics[scale=0.55]{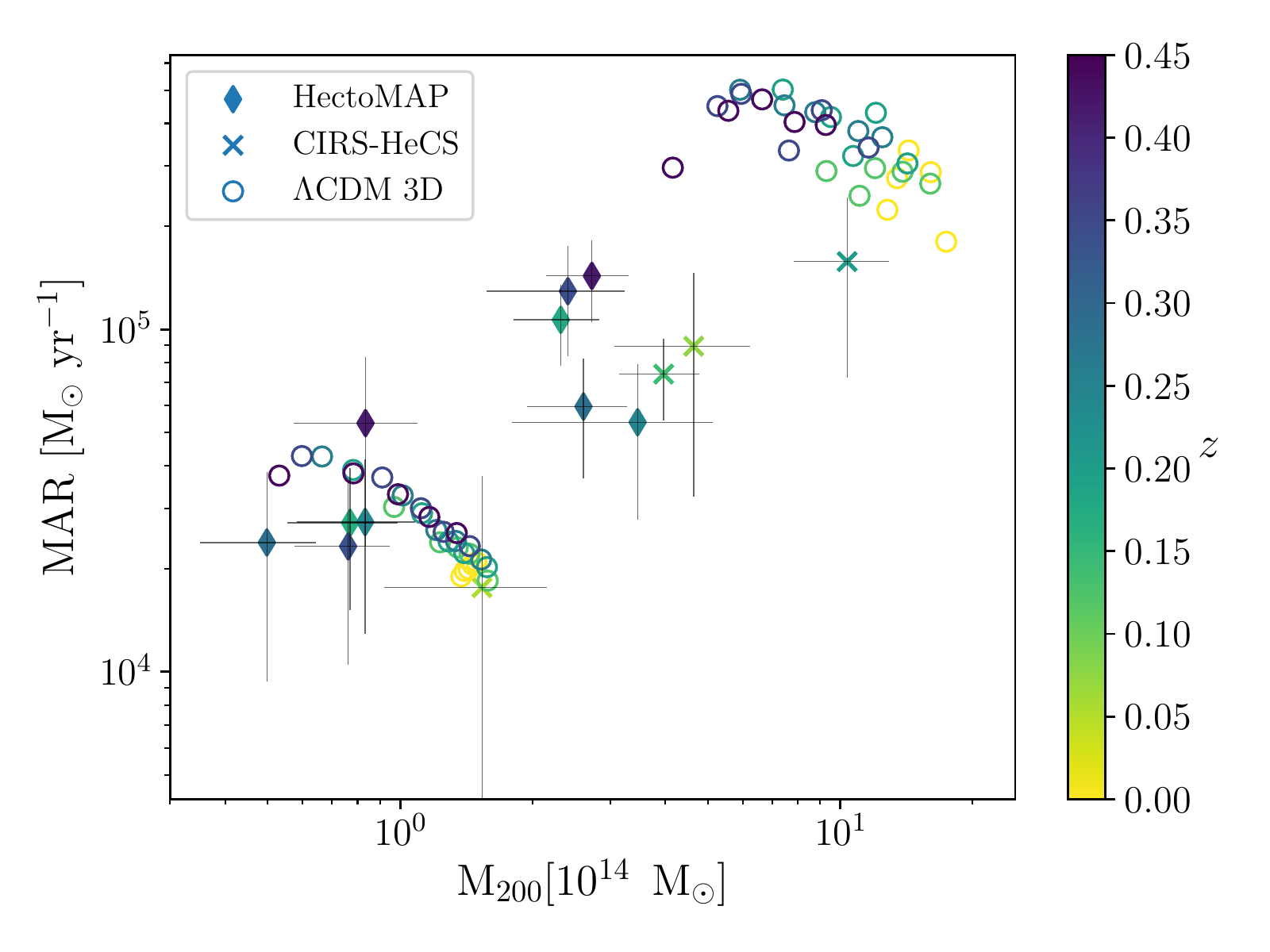}
\caption{MAR of stacked clusters as a function of their mass $M_{200}$, color-coded by redshift. The solid diamonds indicate the 10 stacked clusters from HectoMAP and the crosses show the four stacked clusters from \citet{pizzardo2020} based on observed CIRS and HeCS clusters. The open circles show the median MARs of the simulated clusters, computed from their three-dimensional mass profiles. The simulated data appear separated into two distinct groups because of the two sets of halos considered in the simulations, with masses roughly in the range $(0.6-1.6)\cdot 10^{14} M_{\odot}$ and in the range $(4-16)\cdot 10^{14} M_{\odot}$, respectively. The MAR increases with decreasing mass and increasing redshift, as indicated by the color code and as explained in the text. }\label{marComp} 
\end{center}
\end{figure}

The solid diamonds in Fig. \ref{marComp}  show the resulting MARs of the 10 stacked HectoMAP clusters. Table \ref{rm200_mar_stk} lists the MARs. 
For comparison, the four crosses with error bars show the MARs of the four stacked clusters \citet{pizzardo2020} constructed from the CIRS and HeCS  surveys with the same stacking method we apply to  HectoMAP. To compare the observations with the simulations, we separate each of the 12 simulated samples into five mass bins. We then have 30 low-mass and 30 high-mass subsamples of halos. For each of these subsamples we compute the median of all of its individual halos MARs obtained by applying the recipe for the MAR to the true mass profiles. Fig. \ref{marComp} (open circles) shows the results. 

The MARs of the stacked clusters from HectoMAP are fully consistent with the MARs from the independent sample of stacked clusters from the CIRS and HeCS surveys. The MARs also appear consistent with  $\Lambda$CDM expectations. In particular, the results underscore the positive correlation between the MAR and mass at fixed redshift  and  between the MAR and redshift at fixed mass \citep{pizzardo2020}.

The uncertainties in the HectoMAP MAR estimates differ from the uncertainties for the  CIRS and HeCS results. The mean relative uncertainty for the HectoMAP estimates is $\sim 44\%$; the CIRS-HeCS uncertainty is typically $\sim 64\%$. This difference originates from the difference in the number of galaxies within the $R-v_{\rm los}$ plane of the stacked clusters: there are $\sim 2000-6000$ galaxies for HectoMAP (Table \ref{stkBins}) and $\sim 12000-27000$ for CIRS-HeCS \citep[see the `Total' column of Table 7 in][]{pizzardo2020}. \citet{pizzardo2020} explain this effect. 
For a  larger number of galaxies within the $R-v_{\rm los}$ plane of the stacked cluster, the density contrast between the galaxy cluster members and the fore- and background is smaller.
The distribution of foreground and background galaxies also tends to be more uniform. These effects  increase the uncertainty in the caustic mass profile because the boundary between the cluster and its surroundings is less well defined. Consequently the MAR is also more uncertain. However, the larger number density in the $R-v_{\rm los}$ plane affects only the uncertainties in the mass profiles. The mass profiles themselves and the MARs are substantially unaffected. In other words, the MAR estimates based on the caustic mass profiles of stacked clusters are robust. 

\citet{pizzardo2020} argue that the robust correlations among the MAR, $M_{200}$, and redshift are linked directly to the correlations between $M_{200}$, redshift, and  the mass of the infalling shell with  fixed radial thickness. Unlike the MAR, the mass of the infalling shell is unrelated to $v_{\rm inf}$. 
In the analysis of \citet{pizzardo2020}, the spherical infall model returns an average thickness $\delta_s R_i \approx 0.5R_{200}$. Thus in addition to considering the individual values of the thickness of the infalling shells,  \citet{pizzardo2020}  also consider the mass $M_{2-2.5}$ of the spherical shell with inner and outer radii $2R_{200}$ and $2.5R_{200}$. 

\begin{figure}
\begin{center}
\includegraphics[scale=0.55]{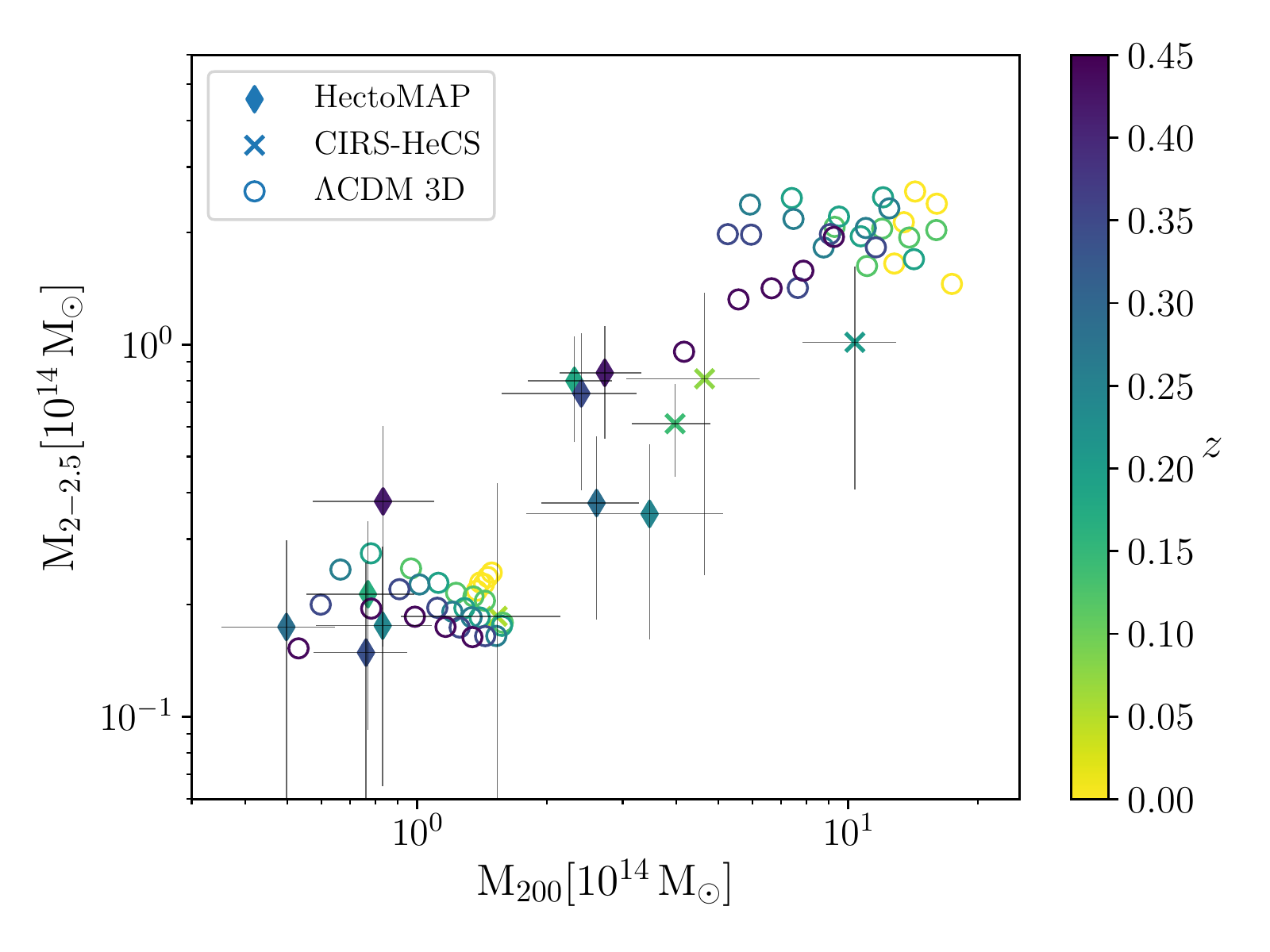} 
\caption{$M_{2-2.5}$ of the stacked clusters as a function of their mass $M_{200}$, color-coded by redshift. The symbols are as in Fig. \ref{marComp}.}\label{mshellComp}
\end{center}
\end{figure}

We also compute $M_{2-2.5}$ for each HectoMAP stacked cluster. We compare the results with the CIRS and HeCS stacked clusters and with the $\Lambda$CDM predictions. Figure \ref{mshellComp} shows the relation between $M_{2-2.5}$, $M_{200}$, and redshift.
As in  Fig. \ref{marComp}, the $M_{2-2.5}$s for the HectoMAP stacked clusters  are consistent with those from CIRS and HeCS and with the estimates based on  true mass profiles from $\Lambda$CDM 
simulations. 

Comparison between Fig. \ref{mshellComp} and Fig. \ref{marComp} confirms the correlation between MAR and $M_{2-2.5}$ for  $z\lesssim 0.3$, but  $M_{2-2.5}$ cannot be used as a proxy for the MAR at higher redshift.
Strikingly, the simulated data (empty circles) in Fig. \ref{mshellComp} have a  distribution that differs from the one in Fig. \ref{marComp} for the  $z\sim 0.35-0.44$ range that exceeds the largest redshifts investigated by \citet{pizzardo2020}. At 
these larger redshifts, the clear increase of the MAR with increasing redshift at fixed mass is absent for $M_{2-2.5}$. The dependence of the MAR on redshift revealed by using shells with increasing width for larger redshifts is an expected feature of  hierarchical structure formation. Starting with two halos of the same mass at different redshifts, the halo at greater redshift  grows faster than the halo at lower redshift; this difference occurs because the higher-redshift halo is embedded within a higher-overdensity region. 
This result indicates that, in general, the radial velocity of the infalling region cannot be ignored in estimating the MAR. 

\begin{figure}
\begin{center}
\includegraphics[scale=0.55]{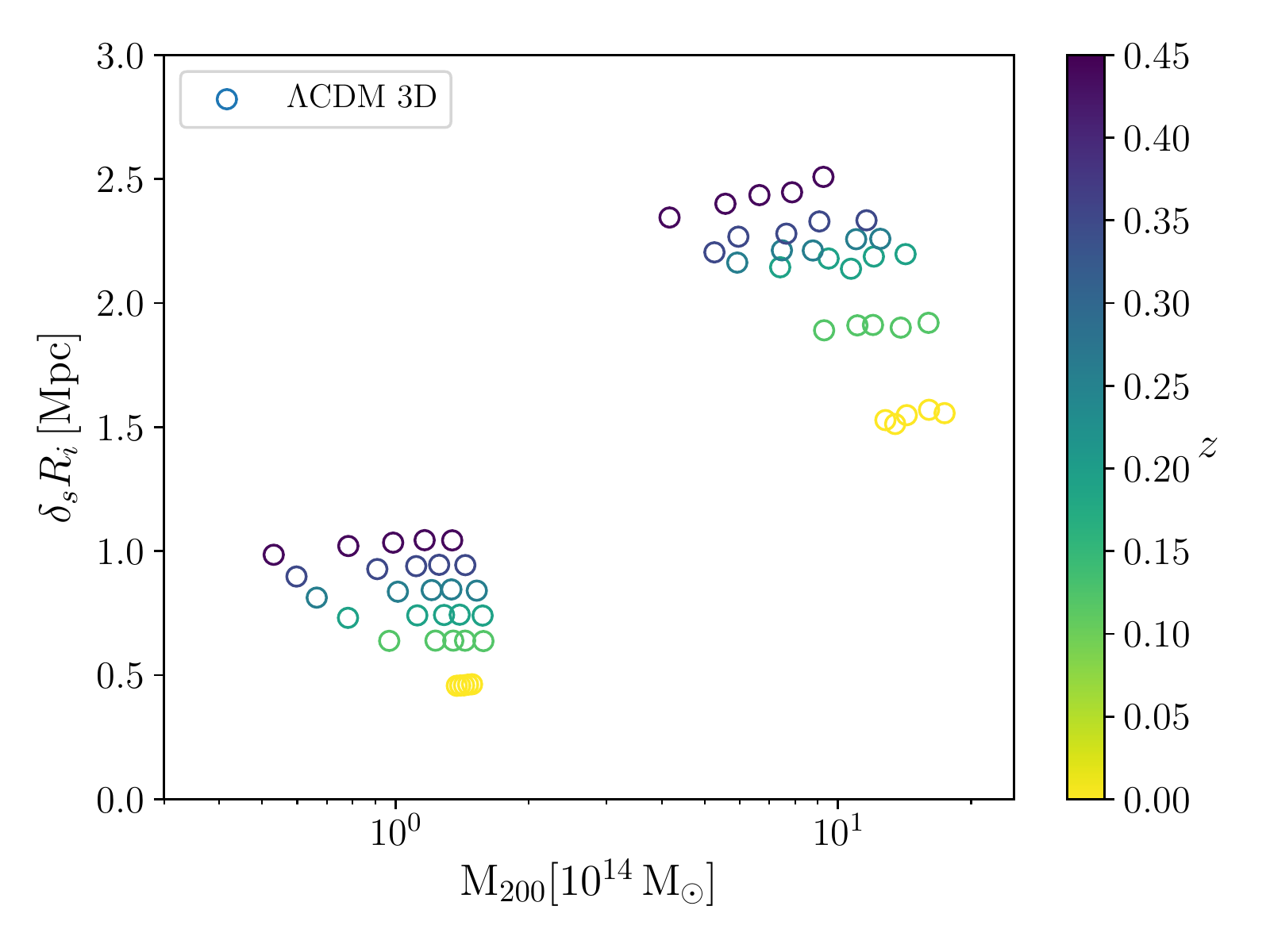} 
\caption{Thickness of the infalling shell of the simulated halos as a function of their mass $M_{200}$, color coded by redshift.}\label{thickness}
\end{center}
\end{figure}

To underscore this point, Fig. \ref{thickness} shows the thickness of the infalling shell, $\delta_s R_i$, as a function of halo redshift and mass. The thickness, $\delta_s R_i$, increases by $\sim 200\%$ and $\sim 75\%$ from $z=0$ to $z=0.44$, for low- and high-mass halos, respectively. At fixed mass, the shell thickness depends mainly on the infall velocity.  Fig. \ref{thickness} thus shows that the absolute value of $v_{\rm inf}$ steadily increases with redshift. The radial velocity profiles of halos confirm this behavior. This behavior of $v_{\rm inf}$ also explains why the correlation between MAR and $M_{2-2.5}$ weakens as the redshift increases. 
The MAR increases with redshift because the infalling shells of high-redshift halos are consistently thicker than those for the low-redshift counterparts. 

\section{Discussion}
\label{sec:discussion}

The independent sample of HectoMAP clusters reproduces the observed MARs of the HeCS and CIRS clusters \citep{pizzardo2020}. The HectoMAP clusters extend the redshift range of  observed MARs to $z\sim 0.42$. These observed MARs are consistent with $\Lambda$CDM expectations. 

Here we discuss the potential for reducing the error in the MAR and thus increasing its power to constrain models for structure formation (Sect. \ref{sec:disc-obs}). We also highlight basic model predictions for higher accretion at greater redshift. Observations will soon access these redshifts, again increasing in the power of the MAR as a discriminant among models (Sect. \ref{sec:disc-theo}). Future approaches to computing the MAR will also benefit from enhanced hydrodynamical simulations covering large
volumes (Sect. \ref{sec:disc-theo}).

\subsection{Observational Challenges}\label{sec:disc-obs}

Combining weak gravitational lensing with the caustic technique holds promise for reducing the uncertainty in the MAR estimates. This ambitious project requires deep and dense spectroscopic surveys along with extensive imaging surveys. Imaging surveys already exist and dense spectroscopic surveys of clusters over a large redshift range will be possible with, e.g., the Prime Focus Spectrograph (PFS) on Subaru \citep{Tamura16}.

Deep, dense spectroscopic surveys of clusters can substantially reduce both systematic and statistical errors affecting weak-lensing mass estimates. The spectroscopic surveys can robustly identify potentially contaminating foreground and background structures, thus enabling correction for their effects.  Very large, deep spectroscopic surveys also promise to limit the systematic error in the source redshift currently based on photometric redshifts \citep{vonderLinden14}. 

At low redshifts, $\lesssim 0.1$, spectroscopic surveys of massive clusters and the weakly lensed  background galaxies are feasible. These surveys enable spectroscopic tomographic weak lensing. \citet{dellantonio20}, using spectroscopic redshifts of A2029 \citep{Sohn2019A2029}, at $z=0.078$, extracted the tangential ellipticity of the background galaxies with $z\lesssim0.8$. The weak-lensing mass  agrees with the X-ray mass. This approach avoids calibration issues inherent in the use of photometric redshifts for background sources, and it effectively removes contamination of the lensing signal by faint cluster members.

The signal-to-noise (S/N)  of this method is currently limited by  small samples of redshifts for faint galaxies. Powerful spectrographs such as the PFS on Subaru will provide spectroscopic catalogs of clusters with $\sim 6000$ cluster members and $\sim 15,000$ background objects; such catalogs will allow spectrotomographic weak-lensing measurements with S/Ns comparable to current photometric-redshift-based weak-lensing measurements for hundreds of galaxy clusters \citep{dellantonio20}. 
This method holds  promise for improving the weak-lensing precision of cluster-mass profiles at large radius where the MAR can be derived.

At higher redshift, a synergy between weak-lensing and caustic mass profiles could result in a substantial decrease of the uncertainty in the MARs. With dense spectroscopic redshift catalogs of clusters, the caustic mass profiles are unbiased because projection effects decrease. However, they  have large uncertainties that do not decrease significantly with better sampling. In contrast, weak-lensing mass profiles are increasingly precise  \citep[e.g.,][]{Umetsu11,Umetsu13}, and some of the inherent biases will be reduced with large spectroscopic surveys. Taking advantage of the respective strengths of the two methods promises much more robust, unbiased, and accurate mass profiles extending to a large enough radius for the determination of the MAR.       

Powerful instruments such as the PFS on Subaru \citep{takada14} have the potential to return large sets of members of individual galaxy clusters. These surveys would allow the determination of MARs for individual clusters. For HectoMAP, however,  we must stack the clusters to obtain large samples \citep{Rines2006CIRS,Biviano10,Serra2011,Biviano20,pizzardo2020} and to obtain the most robust possible results (Sect. \ref{sec:ct-profiles}). Stacking does have the advantage that it averages over spatial and kinematic anisotropies. Extensive redshift surveys like DESI \citep{Dey19} will yield large samples of sparsely sampled clusters, potentially yielding stacked systems similar to those we construct from the HectoMAP. All of these future surveys will access systems at higher redshift.

\subsection{Theoretical Predictions and Challenges}\label{sec:disc-theo}

Deeper redshift surveys allow estimation of the MAR at higher redshifts where the MAR at fixed cluster mass should be larger \citep{vandenbosch02,mcbride2009,Fakhouri2010,vandenbosch14,Correa15,Diemer2017sparta2}. The HectoMAP sample extends to $z\sim 0.42$. Here we use simulations to briefly explore the MAR at higher redshift. 

\begin{figure}
\begin{center}
\includegraphics[scale=0.55]{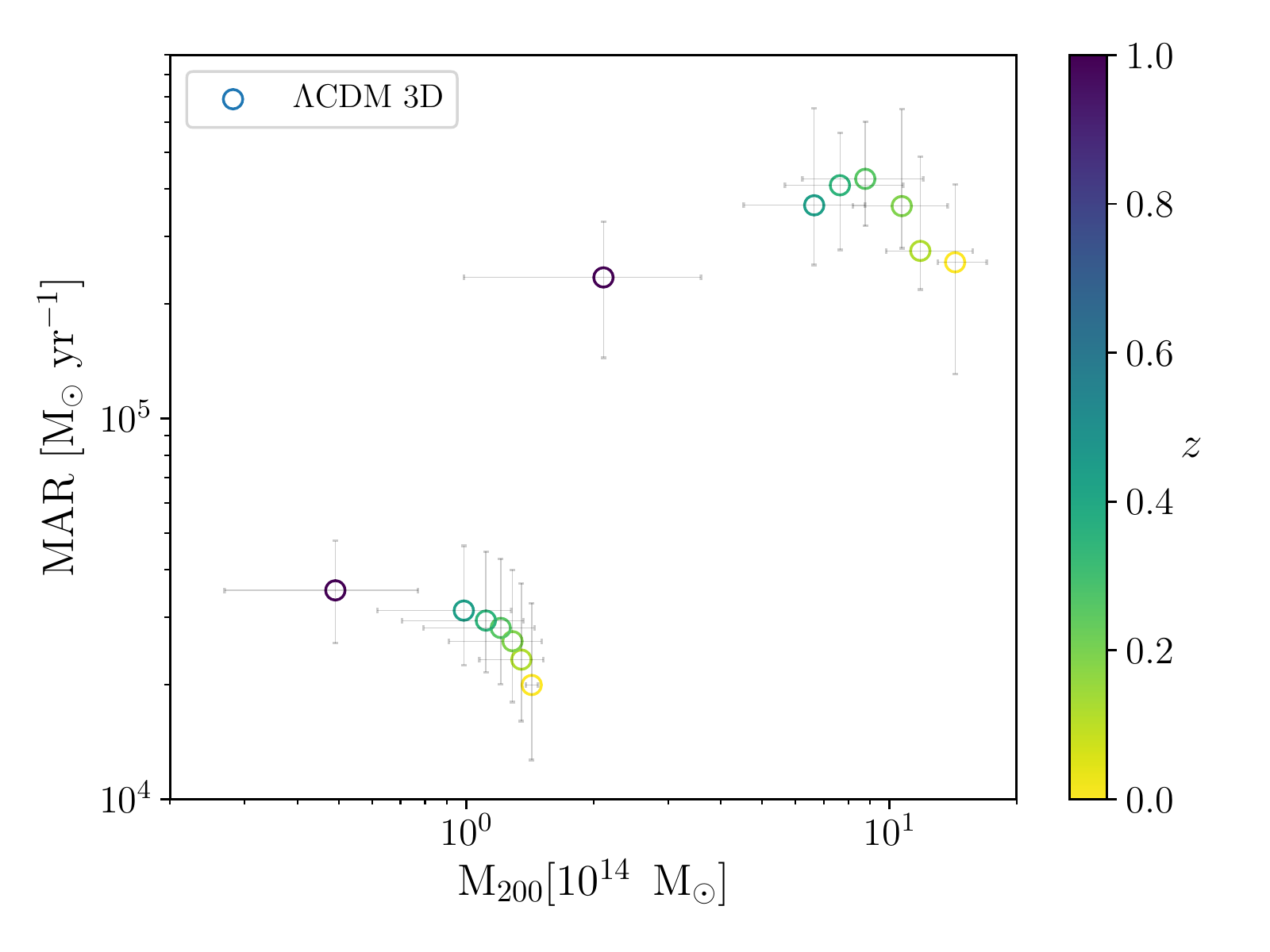}
\caption{MARs of synthetic clusters as a function of  mass $M_{200}$, color-coded by redshift, covering  the range $z=[0,1]$. }\label{marz1} 
\end{center}
\end{figure}

Fig. \ref{marz1} shows the MARs of synthetic clusters in the redshift range $z=[0,1]$. In the low- and high-mass samples (bottom left and top right, respectively), the six circles at lower redshifts, $0-0.44$, use the same simulation data as  in  Fig. \ref{marComp} (Sect. \ref{sec:mar-estimates}), but at each redshift we use only one mass bin for each sample. The upper and lower dark blue circles indicate the median MARs as a function of the median $M_{200}$ of the progenitors at $z=1$ of the halos in the high- and low-mass bin, respectively. The error bars show the $68$th percentile ranges of mass and MAR. The MARs of high-redshift halos are much greater than the MARs of present epoch halos with similar mass:  at $z=1$ the progenitors of the high-mass halos have masses $\sim (1-4)\cdot 10^{14} M_\odot$. This mass range overlaps the low-mass bin at lower redshifts. However, the MARs of the $z=1$ halos are an order of magnitude larger than their counterparts at $z=0$.  

At higher redshift, the MAR may be more sensitive to the nature of dark matter, dark energy, and gravity \citep{Cimatti08,Guzzo08,Baldi2012CoDECS,Candlish16}. New-generation simulations including the hydrodynamical $N$-body Illustris TNG simulation suite \citep{Springel10,Weinberger17,Pillepich18} are a crucial platform for exploring the entire range of physical insights possible with the determination of the MAR at greater redshift. 

The Illustris TNG data release \citep{Nelson19} provides catalogs of galaxies for cleaner comparison with the data. These galaxy catalogs enable the computation of the MAR of simulated clusters based on galaxies rather than the underlying dark matter field.  
The comparison between the models and the data includes the following improvements: (i) simulated galaxies are a better proxy for observed galaxies, (ii) matter \citep{Springel18} and velocity \citep{Ye17,Kuruvilla20} biases that could affect dark-matter-only estimators are minimized, and (iii) the center of a cluster can be identified with a bright galaxy \citep{Sandage73,Dressler79,Dubinski98} or by applying the center-of-light method to cluster member galaxies \citep{Robotham11} in analogy with observed systems.

Larger simulated volumes will permit increases in the statistical significance of theoretical constraints. Large volumes are important for obtaining adequate samples of clusters at the upper end of the mass function. These massive systems place tight constraints on structure formation models \citep{BBKS86,White02,Courtin11,Cui12,giocoli2018weak}. These systems are also the candidate systems most likely to be accessible observationally. 

\section{Conclusion}
\label{sec:conclusion}

We derive MARs that characterize 321 clusters in the HectoMAP Cluster Survey \citep{sohn2021cluster}. The cluster sample covers the redshift range $0.17\lesssim z\lesssim 0.42$ and mass range $M_{200}\approx (0.5 - 3.5)\cdot 10^{14} M_\odot$.

To estimate the MARs, we adopt the approach of \citet{pizzardo2020}. They apply the caustic technique \citep{Diaferio99,Serra2011} to estimate the mass profiles of the clusters at $\sim 2-3 R_{200}$, where accretion occurs. They adopt a spherical accretion prescription \citep{deBoni2016} to evaluate the MAR. $N$-body simulations show that this recipe returns MARs within $19\%$ of the MARs obtained by applying the same recipe to the true mass profiles of synthetic clusters.  The technique is robust against the typical photometric and spectroscopic incompleteness of spectroscopic redshift surveys \citep{pizzardo2020}.

The HectoMAP cluster MARs  agree well with the MARs derived \citep{pizzardo2020} for independent samples of lower-redshift clusters, CIRS \citep{Rines2006CIRS} and HeCS \citep{Rines2013HeCS} at $z \lesssim 0.3$. In the low-velocity dispersion bin, with $M_{200}\sim 0.7\cdot 10^{14} M_\odot$, the average MAR is $\sim 3\cdot 10^4 M_\odot$~yr$^{-1}$, whereas in the high-velocity dispersion bin, with $M_{200}\sim 2.8\cdot 10^{14} M_\odot$, the average MAR is $\sim 1\cdot 10^5 M_\odot$~yr$^{-1}$. The uncertainty in the HectoMAP cluster MAR measurements is $\sim 44\%$, roughly one-third smaller than that of \citet{pizzardo2020}.
We show that an increase in the width of the mass shell with redshift is critical for obtaining the correct MAR at larger redshift.

We compare the HectoMAP MARs  with  $\Lambda$CDM  using the L-CoDECS $N$-body simulation suite \citep{Baldi2012CoDECS}. We apply the MAR recipe to  true mass profiles of six samples of 2000 synthetic clusters each with mass $M_{200}\sim 1.43\cdot 10^{14} M_\odot$, and of six samples of 50 synthetic clusters each with mass $M_{200}\sim 1.43\cdot 10^{15} M_\odot$; we sample the two mass groups at different redshifts in the range $0\leq z\leq 0.44$. The MARs of 10 HectoMAP stacked clusters increase with mass at fixed redshift and with redshift at fixed mass as expected in $\Lambda$CDM. These MARs agree with the MARs estimated in $N$-body simulations \citep{vandenbosch02,mcbride2009,Fakhouri2010,vandenbosch14,Diemer2017sparta2}.

Larger cluster catalogs with more densely sampled systems reaching greater redshift and a more extensive cluster-mass range will provide more sensitive measures of the MAR. The dependence of the MAR on both redshift and cluster mass can provide new insights into the development of structure in the universe. Higher-redshift catalogs extend the MAR probes to epochs where the mass accretion is larger and the sensitivity to the details of the $\Lambda$CDM paradigm is more pronounced \citep{Cimatti08,Guzzo08,Baldi2012CoDECS,Candlish16}. These measures can complement approaches based on large-scale statistical analyses. 

Large dense surveys obtained with heavily multiplexed spectrographs on large telescopes \citep[e.g. the PFS on Subaru; see][]{Tamura16} are crucial for reducing the uncertainty in the measurement of the MAR. The exploitation of weak-gravitational-lensing mass profiles estimated with sophisticated mass reconstruction methods \citep[e.g.,][]{Umetsu11,Umetsu13} combined with caustic profiles will improve the accuracy of mass profiles at large distances from the cluster center.

Finally, the new generation of hydrodynamical simulations like the Illustris TNG suite \citep{Springel10,Weinberger17,Pillepich18}  improves theoretical constraints on cluster accretion, including the high-redshift regime. These simulations  allow the computation of the MAR directly from galaxies, thus improving consistency with results from observations. Larger simulated volumes will allow the identification of larger sets of the most massive halos, paving the way to tighter testing of the structure formation model.

\begin{acknowledgements}
We thank the referee for suggestions that improved the clarity of the paper.
We thank Marco Baldi for providing snapshots of the $\Lambda$CDM L-CoDECS $N$-body simulation. We acknowledge Benedikt Diemer for useful discussions.
The graduate-student fellowship of Michele Pizzardo is supported by the Italian Ministry of Education, University and Research (MIUR) under the Departments of Excellence grant L.232/2016. We also acknowledge partial support from the INFN grant InDark. The Smithsonian Institution supports the research of Margaret Geller and Jubee Sohn. This research has made use of NASA’s Astrophysics Data System Bibliographic Services.
\end{acknowledgements}

\bibliography{michele.bib}{}
\bibliographystyle{aasjournal}

\end{document}